\documentclass{PoS}

\usepackage{xspace}
\newcommand{\hess}{H.E.S.S.\xspace}
\newcommand{\fermi}{\textit{Fermi}-LAT\xspace}
\newcommand{\phasetwo}{\hess~II\xspace}
\newcommand{\phaseone}{\hess~I\xspace}

\title{Extragalactic Observations with HESS: Past and Future}

\ShortTitle{Extragalactic Observations with H.E.S.S.}

\author{\speaker{Andrew~M.~Taylor on behalf of the H.E.S.S. collaboration}\\
        Dublin Institute for Advanced Studies, 31 Fitzwilliam Place, Dublin 2, Ireland\\
        E-mail: \email{taylora@cp.dias.ie}}

\author{David Sanchez\\
        Laboratoire d'Annecy-le-Vieux de Physique des Particules, Universit\'{e} Savoie Mont-Blanc, CNRS/IN2P3, F-74941 Annecy-le-Vieux, France\\
        E-mail: \email{david.sanchez@lapp.in2p3.fr}}

\author{Matteo Cerruti\\
        Laboratoire de Physique Nucl\'eaire et de Hautes Energies (LPNHE), 4 place Jussieu, F-75252, Paris Cedex 5, France\\
        E-mail: \email{mcerruti@lpnhe.in2p3.fr}}

\abstract{The present \phasetwo era of the \hess collaboration follows from the 
successful upgrade of the \hess array, and the first published results 
obtained with this new instrument. Thanks to these achievements, a 
lower energy threshold with \phasetwo than that obtained previously with 
\phaseone has been clearly demonstrated. The success of these developments has now 
opened up a whole new lower-energy window to the Universe. I here 
explore, from the extragalactic point of view, both what has been learnt 
so far from the \hess observations through this new window, and postulate 
also on what else might still be seen beyond this. I will firstly reflect 
upon the AGN result highlights, obtained using the new \phasetwo instrument. 
Specifically, the rise in detections of bright FSRQ AGN with \phasetwo 
will be addressed [eg. PKS 0736 (z$\approx $0.19), PKS 1510 (z$\approx $0.36), and 3C 279 (z$\approx$ 0.54)]. 
Hand in hand with this recent progress, the benefits brought to our efforts to
observe GRBs through both access to this new lower energy window, and 
the quick instrument response to ToO alerts, will be covered. 
Furthermore, the potential for the discovery of new transient phenomena in the
\phasetwo era will also be emphasised. Lastly, the
question as to lessons learnt from previous \phaseone AGN results about
the intrinsic source spectra, primarily focusing on the HBL class, will 
be considered.}

\FullConference{35th International Cosmic Ray Conference – ICRC217-\\
		10-20 July, 2017\\
		Bexco, Busan, Korea}

\begin{document}

\section{Introduction}

%Some background on gamma-ray astrophysics history, first sources, + present day status summary
Extragalactic high energy gamma-ray astronomy has developed from an emerging discipline 
into a fully fledged research field over the past several decades.
Following its initial success in the 1990s with the first AGN discoveries of Mrk~421 and Mrk~501 
\cite{Punch:1992xw,Quinn:1996dj}, the field now boasts the detection of more than 
60 AGN at very high energies (VHE) by ground-based gamma-ray instruments~\footnote{See http://tevcat.uchicago.edu for an up-to-date list.}.

%Growth in variety of different AGN classes detected (BL Lacs, FSRQs, radio galaxies)
A consequence of these observational achievements has been a broadening in the array of 
different AGN subclasses, believed to represent various manifestations of a single (few) 
AGN type(s) \cite{Urry:1995mg}, at these energies. These range from the bright beamed blazars 
of both BL Lac and flat spectrum radio quasar (FSRQ) type, the most numerously observed AGN 
subclass at VHE, to their dimmer weakly beamed/unbeamed counterparts, radio galaxies.

%Different aspects of information provided by these (spatial info, temporal info) + complementarity of these different perspectives in light of unification picture
The jet-beamed blazar family members, are observed as point-like sources. For these, 
information about the spatial extent of the emission 
site may be encoded into the temporal structure of the flux that they emit.
Indeed, the most challenging/enlightening results from observations of such temporal structure 
information, come from the most intense outbursts (such as that of PKS~2155-304 
\cite{Aharonian:2007ig} in 2006). Such extreme 
bright episodic emission has lead to tight constraints being placed upon the size of the 
emission region and the jet Doppler factor.

%Present status of IACTs instruments in the field + 2012 upgrades these received
At present, \hess, based in Namibia, is one of the principal stereoscopic Cherenkov telescope 
instruments currently in operation. This sensitive stereoscopic Cherenkov telescope instrument 
provides a unique VHE perspective on the southern hemisphere regions of the sky. The 
achievements of this instrument have played a key part in bringing about the present 
flourishing status of the field. Furthermore, an upgrade of this experiment through
the completed installation of a massive 28~m telescope at the centre of the original array in 
2012, marked the onset of the \phasetwo era. This upgrade resulted in significant improvements 
in the instrument's low energy sensitivity, reducing significantly its threshold 
energy \cite{Holler:2015tca}.

%Discussion on recent addition of monitoring instruments
Recent years have also seen the arrival of new monitoring instruments, 
with FACT \cite{Dorner:2015jka}, and the now completed HAWC-300 
\cite{Pretz:2015zja}, collectively able to provide wide field of view and sensitive
effective AGN monitoring. The complementarity provided by the monitoring and
follow-ups through both the broad sky coverage, and the in-depth low energy threshold 
targeted observations, make promising the prospects for further growth in the 
coming years. Such collaborative efforts allow  what may be obtained 
from this present generation instrument to be maximised before the arrival of the 
next generation CTA north and south instruments \cite{Consortium:2010bc}.

%Outline of contents
In the following, several of the key recent \hess observational developments in AGN 
gamma-ray astrophysics will be covered. In section~\ref{HESSII_Era}, a discussion
on the dawn of the new \phasetwo era will be addressed. Starting in subsection
\ref{Rise_of_FSRQs}, the first \phasetwo era results will be presented, 
noting in particular the rise of the detection of FSRQ type blazars. In 
subsection~\ref{AGN_ToOs} 
current efforts utilising wide field-of-view VHE AGN monitoring instruments as 
efficient trigger alerts will be discussed. Following this, 
subsection~\ref{new_sources} considers potential new sources which may be
detectable within the new \phasetwo era. Lastly, in section~\ref{AGN_Lessons}, a 
summary of the lessons learnt about the intrinsic spectra of AGN (primarily HBL) 
detected by \phaseone will be presented. The conclusions to this discourse will 
be provided in section~\ref{Conclusion}.

\section{HESS II Era}
\label{HESSII_Era}

%{\bf First results....(PKS 2155, PG 1553)}\\
Since October 2012 a fifth telescope at the centre of 
the original \hess array, has been operational, taking data in coordination
with the other \phaseone telescopes. This five telescope set-up is referred to 
as \phasetwo. The analysis of the data taken in this new setup may be
made either utilising the information from all of the telescopes (hybrid),
or utilising only the information from the fifth telescope (mono),
\cite{Holler:2015uca}, each providing niche advantages depending on the
source type.

One of the recent \phasetwo highlights which utilised the hybrid analysis was
the detection of a new extreme HBL candidate, 1RXS~J023832.6-311658.
Little is presently understood about this class of object, which demonstrates
a continuation of its high energy (HE) hard spectral index into the VHE
domain, without evidence for a cutoff. Furthermore, this blazar class appears
to show little evidence of variability in their lightcurves, %REF FOR THIS?
and curious evidence
of correlations found have suggested that preferred directions of these sources 
exist long ``voidy'' lines of sight \cite{Furniss:2014bna}. 
Such limitations in our understanding of this class, in part, is due to 
the small number of such objects so far having been discovered, highlighting the 
need to search for more such objects.

The first \phasetwo AGN results which utilised the mono analysis came from 
observational campaigns taken with this new setup in 2013 and 2014. These
observations were of the HBL blazars PKS~2155-304 and PG~1553+113, which were 
both found to be in quiescent states during the observation periods \cite{Aharonian:2016ria}.
Despite these lowered activity levels, significant achievements in
lowering the threshold energy utilising the new fifth telescope (mono) data, 
with analysis of this data for the PKS~2155-304 and PG~1553+113 observations 
reaching down to new threshold energies of 80~GeV and 110~GeV, respectively.

\subsection{The Rise of the FSRQs}
\label{Rise_of_FSRQs}
\begin{figure}[t]
\begin{center}
\includegraphics[width=0.5\textwidth]{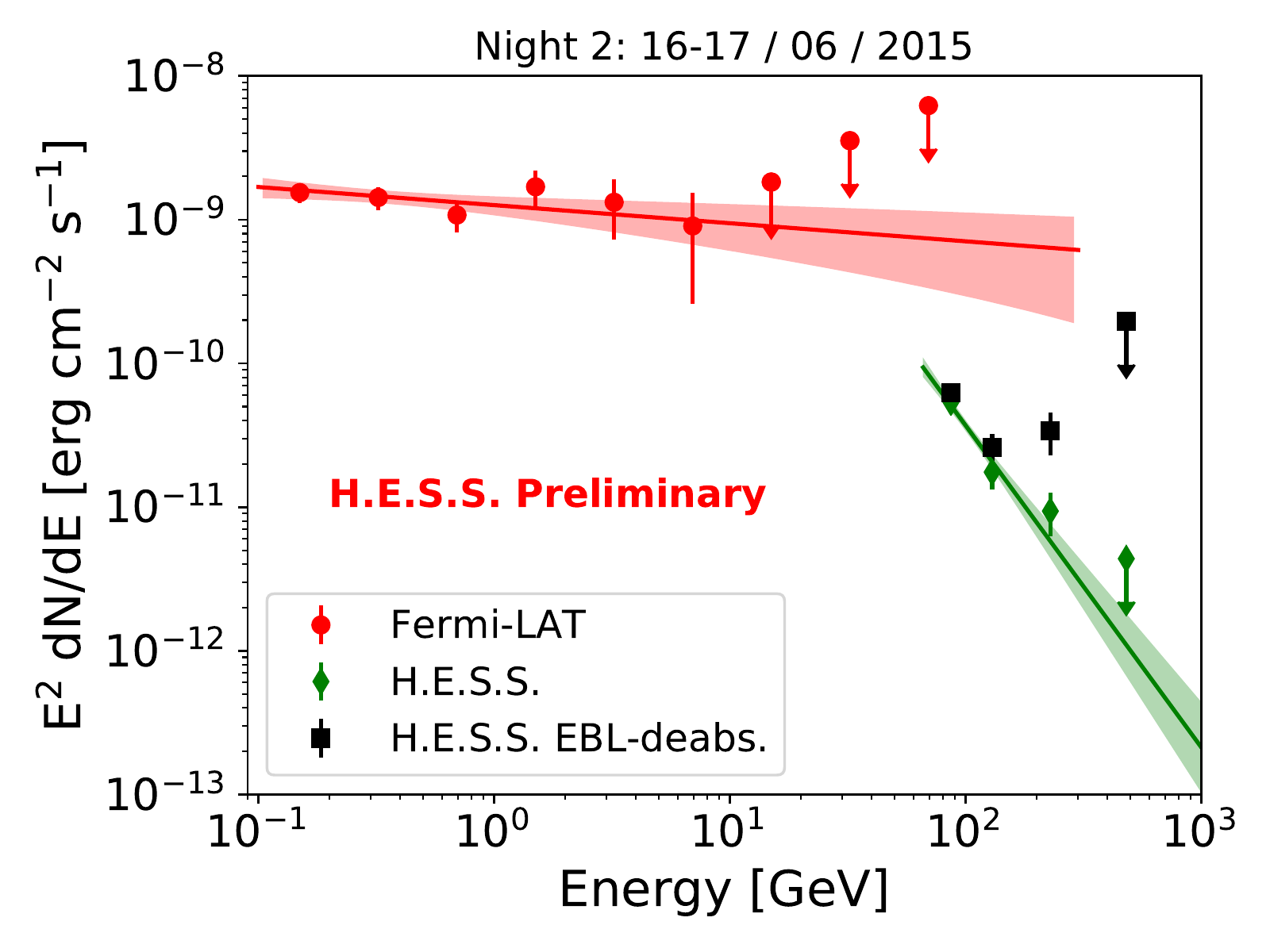}
\caption{Energy flux data points of 3C~279 during its giant outburst in
July 2015.}
\label{3C279_spectra}
\end{center}
\end{figure}

Building on the successful achievement of a lower energy threshold analysis
utilising the \phasetwo, a wider spread of blazar classes (eg. HBLs, LBLs, FSRQs)
and redshifts, has become accessible to the instrument. Indeed, further
proof of the successful lowering of the threshold energy of analysis utilising 
\phasetwo mono data comes from the observations 3C~279, which underwent a
giant outburst back in July 2015. \phasetwo mono analysis of these observations of 
the flare achieved a record low energy threshold energy of 66~GeV for
AGN results with this instrument. % note non-detection of 3C~279 from CT1-4 observations?
Fig.~\ref{3C279_spectra} shows the HESS and Fermi components of its SED of 
during the second night of the flaring outburst.
Along this same vein, the detection of second FSRQ, PKS~0736+017, which underwent 
an outburst in February 2015, was also achieved utilising the analysis of 
\phasetwo data. Again, \phasetwo mono analysis of these observations achieved an energy
threshold of 80~GeV.

Lastly, observations by both \hess and MAGIC of giant flare from another FSRQ, 
namely PKS~1510-089, which underwent an massive outburst in May 2016, collectively
provided exceptional temporal coverage of the flaring event at VHE.

%{\bf Surprises....fast variability + lack of internal absorption}
For all three of these FSRQ flares, the detection of VHE emission from them during
their outbursting episodes presents new surprising and unexpected challenges for 
their modeling. Specifically, the presence of the broad line region (BLR), an 
intense radiation field in the vicinity of the AGN, presents a barrier for the 
escape of VHE emission from within it. In turn, the detection of VHE emission from
these sources can be used to place strong constraints on the position 
of the emission site with respect to the BLR location 
\cite{Tavecchio:2012cs}. With gamma-ray emission beyond 200~GeV detected from
each of these systems during their outbursts, the 
emission site is found to be constrained to sit at a distance beyond 
$r_{\rm BLR}\approx 10^{17}~{\rm cm}\left(L_{\rm disk}/10^{45}~{\rm erg s}^{-1}\right)^{1/2}$ 
\cite{Ghisellini:2009wa}, where $L_{\rm disk}$ is the thermal luminosity of the 
accretion disk.

%{\bf ADD FIGURE HERE TO HIGHLIGHT OPACITY CHALLENGES?}

Information about the spatial size of the emission site, from where the outburst 
originates, is also provided by the minimum temporal variability time-scale
in the observed lightcurve. During their recent outbursts, unexpectedly short 
time-scale variability have been revealed for both objects. Specifically, for 
3C~279, $\sim$minute time-scale structure in the $>$100~MeV HE lightcurves was 
discovered \cite{TheFermi-LAT:2016dss}. Likewise, at VHE, for PKS~1510-089
$\sim$tens of minute time-scale structure. %{\bf NOTE ref. energy for VHE variability}.

Together, both the lack of internal absorption features in the
flaring FSRQ spectra, and the short variability time-scales observed
during the flare, make for rather challenging constraints. Reconciliation
of these two differing results only appears to be possible in a few possible 
scenarios. The first is if the emission site sits sufficiently far out such 
that absorption of the BLR is avoided, potentially allowing the intrinsic 
spectrum to continue as an extrapolation of that in the \fermi domain. The second 
is if the emission originates from only a small subsection of the jet, out 
at distance scales beyond the BLR. 

\subsection{An AGN Transient Machine}
\label{AGN_ToOs}

%{\bf AGN monitoring activities, utilising input from an array}\\
%{\bf of different instruments (ATOM, Swift, Fermi, HAWC, FACT, other IACTs).}
To fully exploit the lowered threshold energy to observe the
larger assortment of VHE AGN in the high redshift Universe demands
efficient wide-field coverage of the transient VHE sky. To this end, full 
advantage is being made of broad multi-wavelength transient event monitoring.
In particular, the \hess collaboration is provided with alerts at optical,
X-ray, HE gamma-ray, and VHE gamma-ray energies. Indeed, for the three AGN
discussed in the previous section, 3C~279, PKS~0736+017, and PKS~1510-089,
\fermi triggers and a \hess trigger during a monitoring campaign, 
instigated the subsequent in-depth observations during their heightened 
activity states.

\begin{figure}[t]
\begin{center}
\includegraphics[width=0.5\textwidth]{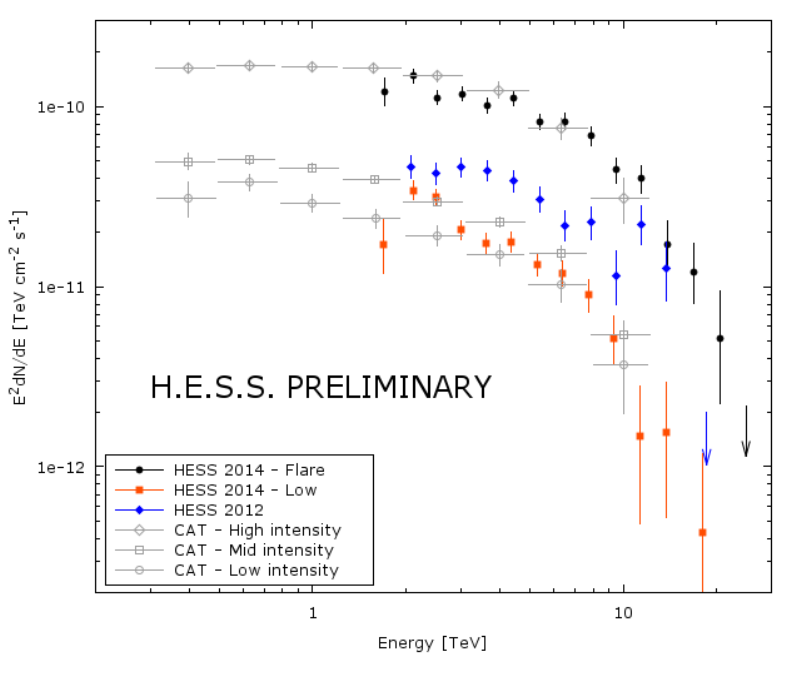}
\caption{The observed spectral data points of Mrk~501 above 300~GeV, from 
observations taken both before and during its extremely bright outburst in 
June 2014 \cite{Cologna:2015mia}.}
\label{Mrk501_spectra}
\end{center}
\end{figure}

%{\bf Mention Mrk 501 2014 flare- FACT alert}
A further example of such effective transient observations is provided by
the \hess observations of Mrk~501 in June 2014. These observations were triggered by 
FACT, an imaging air Cherenkov telescope (IACT) which regularly monitors the activity 
of known VHE AGN, providing alerts for follow-up observation to the VHE gamma-ray 
community. During a giant outburst in 2014, the flux level of Mrk~501 
observed by \hess matched that of the record level, observed by HEGRA back 
in 1997 \cite{DjannatiAtai:1999af}. The obtained spectra both during this flare, 
and in the quiescent state, are shown in fig.~\ref{Mrk501_spectra}.
Of particular note from these results is that the spectrum observed during this 
flare, once extragalactic background light (EBL) absorption had been accounted 
for, showed no signs of a cutoff, continuing as a hard spectrum up to the highest 
energy data point ($\sim 20~$TeV).

\subsection{New VHE Transient Sources}
\label{new_sources}

Alongside the broadened AGN discovery potential which the onset of the 
\phasetwo era has opened up, the possibility to catch new VHE phenomena has 
also been increased. Indeed, a consideration of the competing sensitivities of
\fermi and next generation IACT \cite{Funk:2012ca}, highlights that within the 
overlapping energy region between such instruments, it is the temporal domain 
of the next generation instruments in which the discovery frontier lies.
 
To pursue an exploitation of \phasetwo in this direction, for the catching of 
new transients, a rapid repointing system for the telescopes has been implemented. 
This system has been designed to respond as fast as possible, without human 
intervention, to targets of opportunity (ToOs).

\begin{figure}[t]
\begin{center}
\includegraphics[width=0.5\textwidth]{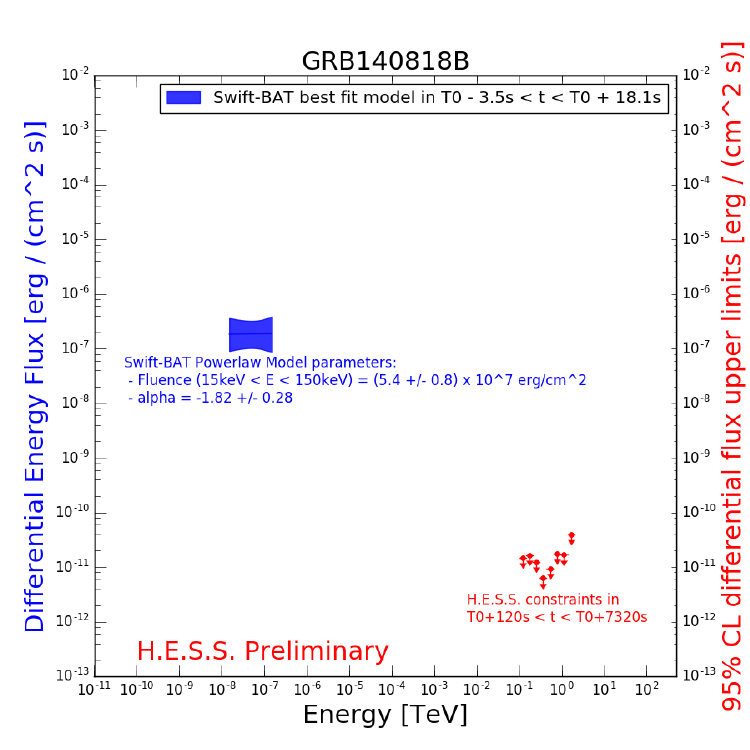}
\caption{Upper limit results obtained from the H.E.S.S. observations for GRB140818B (shown in red). The best fit spectral model for the Fermi-GBM detection is shown in blue.}
\label{GRB140818B_upperlimits}
\end{center}
\end{figure}

%{\bf Discussion on GRB efforts + development of rapid response follow-ups}\\
With an expected fall-off in intensity of the gamma-ray flux from GRB observed 
following the prompt phase emission, motivation exists for rapid response 
follow-up observations of such bursts. A minimisation in the repointing time 
for these observations over the last few years has succeeded to reducing the 
average overall response time to a timescale of $\sim$few~minutes. Upper
limits for such a follow-up observation of GRB~140818B are shown in 
fig.~\ref{GRB140818B_upperlimits}.

%{\bf Discussion on FRB follow-up efforts}\\
%{\bf Discussion on Neutrino follow-up efforts}
Beyond the successes of this \phasetwo rapid response GRB ToO activity, efforts are also 
underway within collaboration to utilise \phasetwo to catch other extragalactic VHE gamma-ray 
emitting phenomena. Specifically, attention is here drawn to the VHE neutrino 
and FRB follow-up observations carried out by the collaboration. With the origins
of both these phenomena remaining unclear, though believed in both cases to be 
(in part) extragalactic in origin, great potential exists for fresh insight about the 
emission mechanism to be provided by such follow-ups.

In particular, for the case of VHE neutrino follow-up observations, a strong potential 
link exists between high-energy neutrinos and gamma rays through the possibility that 
both particles are secondary losses of high energy cosmic rays within the region 
in or around their acceleration site. Provided that both particles types are able
to escape from the source region and arrive to Earth, and the transient event overlaps
sufficiently with the observation window, a detectable flux level is expected 
within the \hess energy range.

% ANTARES event follow-up
An example of the recent improvement in response time to neutrino alerts is the 
follow-up observations of the ANTARES neutrino event on the 30th January 2017. 
These observations took placed only 32 seconds after the reconstructed neutrino 
event occurred. Although searches for a gamma-ray counterpart within this data set 
are still ongoing, the success of the automatic response system has already proved 
itself through this considerable reduction in the follow-up delay time to neutrino
event alerts.

\section{Intrinsic AGN Spectra}
\label{AGN_Lessons}

%{\bf Lessons learnt to date about intrinsic source AGN spectra.}
In parallel with efforts to broaden and deepen the range of sources detected in 
the \phasetwo era, investigations are also being made to consolidate what 
lessons have been learnt from the detection of blazars in the \phaseone era.
Specifically, this has been focused on their intrinsic spectral properties in
this energy range.

The ability to accurately reconstruct the intrinsic spectra of blazars has two limiting
factors. The first of these is dictated by the instrument sensitivity, which dictates the 
photon statistics obtained during observational campaigns of an object. The second of
the constraints comes from the limit in our understanding of the EBL, which at present
is predominantly inferred a mixture of modeling \cite{Franceschini:2008tp} and VHE blazar 
observations \cite{Abramowski:2012ry}.

Following the adoption of the recent EBL model and its associated uncertainty %\cite{} add recent HESS pub here when available 
archival \phaseone data has been recently reviewed. An investigation was carried 
out to collectively infer what the full \phaseone data set revealed about the intrinsic AGN
spectra. In particular, following the assumption that the intrinsic spectra are describable
by power-law or log-parabola functions, the constraints on the free-parameters 
of these functions were obtained. Among other things, the results demonstrated 
the fact that only the brightest AGN flares, for which the highest statistics were 
obtained, showed a preference for log-parabola type spectra. The majority of AGN 
spectra, however, found no statistical preference for such a spectra, preferring 
instead the simpler power-law model.

\section{Conclusion}
\label{Conclusion}

The present epoch of \hess extragalactic observations is one of a maturing and
broadening frontier. Although an increase in the number of BL Lac blazar sources 
detected continues, there has also been an evolution of focus. 
This evolution is primarily thanks to recent improvements in the instrumentation 
which have lowered the energy threshold. Hand in hand with these instrumental 
improvements is the implementation of wide field-of-view monitoring and trigger 
follow-up schemes, allowing efficient capture of bright VHE emission follow-up 
observations.

AGN blazar variability is a well known and familiar phenomena. Despite this, however,
the short-time variability results observed from recent FSRQ outbursts at VHE
are challenging. An example case in point is the blazar 3C~279, which underwent a 
giant outburst in June 2015, demonstrated minute-scale variability at GeV energies. 
Such short-time variability at gamma-ray energies approaches the shortest level 
caught from the BL Lac PKS 2155-304. Since considerable internal absorption for FSRQs 
are expected should the emission zone be located within the BLR, both the compactness 
of the emission zone suggested by the short time-scale structure, and the large 
distance from the BLR region, are collectively rather challenging to reconcile.

Beyond new and deeper AGN observations, the new \phasetwo era offers the promise 
for the discovery of new extragalactic VHE sources. Of particular relevance in 
this domain is the need for rapidly response to ToOs. The successful implementation 
of an automatic rapid response system is here demonstrated through the some of the 
first results obtained with this system.

Hand in hand with efforts looking to the future to explore the new range of
phenomena open to \phasetwo, the lessons learnt from \phaseone are considered.
Looking back at the archival set of \phaseone AGN data, a clear message about
what can be learned about the intrinsic spectra is found. The key message
here being that only in brightest of AGN observations could the higher moments
of the intrinsic spectra be probed.

In summary, exciting and increasingly challenging new results have arisen in the 
maturing discipline of \hess AGN observations. The achievement of these results have 
come about both through the lowering of the instrument threshold allowing the 
detection of bright FSRQ flares, and the utilisation of wide field-of-view VHE AGN 
monitoring systems ensuring efficient follow-up observations of bright flares. These 
developments collectively ensure that this observational frontier continues to both 
broaden and deepen our understanding of extragalactic sources, fundamentally providing 
key insights into how these effective particle accelerators operate.

%%%%%%%%%%%%%%%%%%%%%%%%%%%%%%%%%%%%%%%%%%%%%%%%%%%%%%%%%%%%%%%%%%
\section*{Acknowledgements}The support of the Namibian authorities and of the University of Namibia in facilitating the construction and operation of H.E.S.S. is gratefully acknowledged, as is the support by the German Ministry for Education and Research (BMBF), the Max Planck Society, the German Research Foundation (DFG), the French Ministry for Research, the CNRS-IN2P3 and the Astroparticle Interdisciplinary Programme of the CNRS, the U.K. Science and Technology Facilities Council (STFC), the IPNP of the Charles University, the Czech Science Foundation, the Polish Ministry of Science and Higher Education, the South African Department of Science and Technology and National Research Foundation, the University of Namibia, the Innsbruck University, the Austrian Science Fund (FWF), and the Autrian Federal Ministry for Science, Research and Economy, and by the University of Adelaide and the Australian Research Council. We appreciate the excellent work of the technical support staff in Berlin, Durham, Hamburg, Heidelberg, Palaiseau, Paris, Saclay, and in Namibia in the construction and operation of the equipment. This work benefited from services provided by the H.E.S.S. Virtual Organisation, supported by the national resource providers of the EGI Federation.
%%%%%%%%%%%%%%%%%%%%%%%%%%%%%%%%%%%%%%%%%%%%%%%%%%%%%%%%%%%%%%%%%%

\end{document}